\documentclass[a4paper,12pt]{article}

\usepackage[left=2cm,right=2cm,top=2cm,bottom=3cm]{geometry}

\usepackage{amsfonts}
\usepackage{amssymb}
\usepackage{amsmath}
\usepackage{amsthm}
\usepackage{stmaryrd}
\usepackage{amsbsy}
\usepackage{mathrsfs}
\usepackage{mathtools}
\usepackage{comment}
\usepackage{tikz}

\usepackage[linktoc=all,
        hidelinks,
        backref=page,
        pdftex,
        colorlinks=false,
        pdfstartview=FitV,
        linkcolor=ColorLink,
        citecolor=ColorCite,
        urlcolor=ColorURL,
        hyperindex=true,
        hyperfigures=false%
    ]{hyperref}

\usepackage{cite}
\usepackage{graphicx}
\usepackage{tabularx}
\usepackage{fancyhdr}
\usepackage{mdframed}
\usepackage{relsize}
\usepackage{chngpage}
\usepackage{enumitem}
\usepackage{calrsfs}
\usepackage{appendix}
\usepackage{caption}
\usepackage{longtable}
\usepackage{tensor}
\usepackage{tikz}
\usetikzlibrary{positioning}
\usepackage[smalltableaux,centertableaux,centerboxes]{ytableau}

\numberwithin{equation}{section}

\newcommand{\rd}{\mathrm{d}}
\newcommand{\pd}{\partial}

\newcommand{\myfontbackref}[1]{
    \mbox{\small #1}}
\renewcommand*{\backref}[1]{}
\renewcommand*{\backrefalt}[4]{%
\ifcase #1 \myfontbackref{no citations}
    \or \myfontbackref{\!\!(Cited on page #2)}
    \else \myfontbackref{\!\!(Cited on pages #2)}\fi}

\DeclareRobustCommand{\sscdots}{%
  \vbox{%
    \baselineskip=0.125\normalbaselineskip
    \hbox{.\hspace{-0.25mm}.\hspace{-0.25mm}.}%
    \kern-0.2\baselineskip
  }%
}

\begin{document}

\thispagestyle{empty}
\begin{center}

\vspace*{20pt}

\noindent\rule{\textwidth}{0.5pt}

\vspace{10mm}
{\Large \sc A note on non-Lorentzian duality symmetries}
\vspace*{5mm}

\noindent\rule{\textwidth}{0.5pt}

\vspace{40pt}
Josh A. O'Connor${}^{\,a,}$\footnote{FRIA grantee of the Fund for Scientific Research – FNRS, Belgium.}
and Simon Pekar${}^{\,b}$

\vspace{15pt}
\centering
\href{mailto:josh.o'connor@umons.ac.be}{\texttt{josh.o'connor@umons.ac.be}}
\quad
\href{mailto:simon.pekar@polytechnique.edu}{\texttt{simon.pekar@polytechnique.edu}}

\vspace{15pt}
\centering${}^a${\sl \small
Physique de l’Univers, Champs et Gravitation, Universit\'e de Mons\hspace{6mm}\null\\
Place du Parc 20, 7000 Mons, Belgium\hspace{3mm}\null}\\
\vspace{10pt}
\centering${}^b${\sl \small
Centre de Physique Théorique - CPHT, CNRS, École polytechnique,\hspace{6mm}\null\\
Institut Polytechnique de Paris, 91120 Palaiseau, France \hspace{3mm}\null}\\

\vspace{40pt}
{\sc{Abstract}} 
\end{center}

\noindent
We work out non-Lorentzian dual actions for electromagnetism and linearised gravity, both in the Carrollian and Galilean cases.
This is done in the same way as for Lorentzian theories, by first constructing a parent action that reduces to a pair of dual actions.
In the case of Maxwell theory, each pair of dual actions consists of the known `electric' and `magnetic' limits of the original theories, showing that these limits are related by an off-shell electromagnetic duality.
We have obtained dualities between on one hand the non-Lorentzian contractions of linearised gravity in second-order form, and on the other hand the theories one obtains by gauging the corresponding kinematic algebras.
In the Carrollian contraction, these dual actions reproduce the known `electric' and `magnetic' Carrollian theories of gravity, and we find a similar result in the Galilean case.

\newpage


\setcounter{tocdepth}{2}
\tableofcontents

\section{Introduction}

Duality symmetries are transformations between classically equivalent descriptions of the same system, featuring different fields that obey different equations of motion.
A famous example of this is the invariance of Maxwell's theory of electromagnetism in four dimensions in the absence of sources under the rotation of the electric and magnetic fields.
While this transformation is local in terms of the aforementioned fields, it is non-local in terms of the fundamental gauge field $A_\mu$ and the dual gauge field $\tilde{A}_\mu$ since their field strengths are related to each other via Hodge duality: $\pd_{[\mu} A_{\nu]} = \frac12\,\varepsilon_{\mu\nu}{}^{\rho\sigma} \pd_\rho \tilde A_\sigma$\,.
The equations of motion of one theory are exchanged with the Bianchi identities of the other, and this continues to hold for more complicated theories.

The standard method of constructing dual action principles takes as its input an action for a given field and constructs from it a parent action featuring a new dual field which functions as a Lagrange multiplier.
The original action (which depends only on the curl of the original field) is recovered when the equation of motion for the new field is imposed.
However, if we treat the curl of the original field as independent and impose its equation of motion instead, then we obtain a new action that is dual to the original action and which describes the same degrees of freedom.
At the quantum level it was shown in \cite{Fradkin:1984ai} that pairs of dual actions lead to identical partition functions and thus their Feynman rules are the same.

A dual action for gravity was first constructed in \cite{West:2001as,West:2002jj}.
This was later made explicit and generalised to higher dimensions and higher-spin fields in \cite{Boulanger:2003vs} where the authors managed to recover the Curtright action for the dual graviton in arbitrary dimension together with the correct gauge symmetries \cite{Curtright:1980yk,Aulakh:1986cb}.
In four dimensions, the linearised graviton and dual graviton have the same tensor structure: they are symmetric rank-two tensors.
Moreover, these fields are dynamically equivalent and their action principles are identical.

The aim of the present paper is to apply this procedure to dualise  non-Lorentzian theories.
In particular, we consider Carrollian and Galilean theories that emerge when the speed of light approaches zero and infinity, respectively.
There are several motivations behind taking these limits.
On the Galilean side, it is known that Galilean physics accurately describes our physical world where objects typically move much slower than light.
One can view Galilean physics as the leading-order approximation in a non-relativistic expansion, where the expansion parameter is the speed of the system relative to the speed of light.
On the Carrollian side, Carroll symmetry naturally emerges on zero-signature (null) hypersurfaces.
These surfaces may be at a finite distance like the horizon of a black hole \cite{Donnay:2019jiz} or a gravitational wave \cite{Duval:2017els} or at an infinite distance like null infinity \cite{Duval:2014uoa,Ciambelli:2018wre}.
Thus it is expected that Carrollian physics plays a role in holography.

The non-Lorentzian space-times that we consider here, sometimes misleadingly referred to as non-relativistic, respect a non-Lorentzian principle of relativity and symmetry wherein the Lorentz boosts are replaced by their $c \to 0$ or $c \to \infty$ limits: Carroll boosts and Galilei boosts.
Correspondingly, the non-Lorentzian theories living on such backgrounds will be called either Carroll-invariant or Galilei-invariant if they respect these symmetries.
Examples of such theories include scalar fields, electromagnetism, gravity, higher-form and higher-spin theories \cite{LeBellac:1973unm,Henneaux:1979vn,Dautcourt1990OnTN,Duval:2014uoa,Henneaux:2021yzg,deBoer:2021jej}.
Further examples can be found in string theory \cite{Gomis:2000bd,Sundborg:2000wp} and holography \cite{Taylor:2008tg,Harmark:2017rpg,Ciambelli:2018wre}.
Remarkably, there is evidence to suggest that non-relativistic strings form a solvable sub-sector of string theory, and moreover it is hoped that the large-$c$ and/or small-$c$ expansions of sophisticated theories like string theory or supergravity might be easier to describe and quantise than their Lorentzian counterparts, while still capturing many interesting and crucial physical features.
Thus there is ample motivation for studying these non-Lorentzian limits.

Carrollian and Galilean theories are often defined as the $c\to 0$ and $c \to \infty$ contractions of relativistic theories.
However, there is typically more than one way of defining these limits as one can start from a first-order or second-order formulation of a relativistic theory.\footnote{One can even work at the level of equations of motion without referring to an action principle. This is the way in which Galilean electrodynamics was originally obtained in \cite{LeBellac:1973unm}.}
One of the earliest indications of the existence of multiple contractions are the various Galilean and Carrollian limits of source-free Maxwell theory in four dimensions \cite{Duval:2014uoa}.
They exhibit regimes where the electric field dominates over the magnetic field and vice versa.
In \cite{Henneaux:2021yzg}, it was found that the Carrollian contraction of Lorentz-invariant theories usually gives rise to two possible limits, referred to as `electric' and `magnetic' in direct analogy with the work of Le Bellac and L\'evy-Leblond, although no notion of electromagnetic duality seems to be attached.
In fact, the two limits seemingly differ as one is second-order in time derivatives and features no spatial gradients, while the other is first-order in time derivatives and features spatial gradients.

It was found that the electric and magnetic Carrollian contractions of Maxwell's theory are equivalent \cite{Duval:2014uoa}, while the same was previously known to be true in the Galilean case \cite{LeBellac:1973unm}.
Indeed, the exchange of the electric and magnetic fields in the equations of motion maps each theory into the other.
This was an early example of a non-Lorentzian duality symmetry, and it is naturally inherited from the usual relativistic duality symmetry.
More recently \cite{Bekaert:2024itn}, the two Carrollian limits of a conformally coupled scalar field were also mapped into one another through a non-local transformation.
In this reference, it was suggested that this duality could be extended to arbitrary spin.
This begs the question: are the various electric and magnetic limits defined in the non-Lorentzian literature equivalent?
In other words, do there exist non-Lorentzian duality symmetries?
In this paper, we answer positively by constructing non-Lorentzian parent actions for both Carrollian and Galilean field theories, including Maxwell's theory in four dimensions, $D$-dimensional $p$-form electrodynamics, and four-dimensional linearised gravity.

This paper is organised as follows.
In Section~\ref{sec:Maxwell}, we build parent actions and dual actions for Carrollian and Galilean Maxwell theory in four dimensions.
This is generalised in Section~\ref{sec:p-form} to $p$-form electrodynamics in $D$ dimensions.
In Section~\ref{sec:Fierz-Pauli}, we solve the same problem for linearised gravity in first-order form, thus relating to each other the `electric' and `magnetic' Carrollian theories of gravity, and we obtain an analogous result for the Galilean case.
We conclude in Section~\ref{sec:conclusions} with a discussion of our results and an outlook for future lines of research.

\section{Maxwell theory in four dimensions} \label{sec:Maxwell}

In this Section, we revisit the early study of Carrollian and Galilean electrodynamics of \cite{LeBellac:1973unm,Duval:2014uoa} (see also \cite{Henneaux:2021yzg,Bergshoeff:2022qkx}) and show that the `electric' and `magnetic' limits in both cases are related by an off-shell electromagnetic duality.

To begin we will consider the relativistic Maxwell action
\begin{align}\label{eq:action_Maxwell}
    S_\text{M}[A]=\int\!\rd^4x\,\Big({-\tfrac14}F[A]_{\mu\nu}F[A]^{\mu\nu}\Big)\,,
\end{align}
where $F[A]_{\mu\nu}:=2\,\pd_{[\mu}A_{\nu]}$ is the field strength of the Maxwell field that is invariant under the usual gauge symmetry $\delta A_\mu=\pd_\mu\lambda$\,.
We remind the reader that in four dimensions this action is dual to an identical action featuring a dual Maxwell field $\tilde{A}_\mu$\,.
In order to show this duality we introduce the parent action
\begin{align}\label{eq:parent_Maxwell}
    \hat{S}_{\text{M}}[F,\tilde{A}]=\int\!\rd^4x\,\Big({-\tfrac14}F_{\mu\nu}F^{\mu\nu}+\tfrac12\,\varepsilon^{\mu\nu\rho\sigma}F_{\mu\nu}\pd_\rho\tilde{A}_\sigma\Big)\,.
\end{align}
There are now two independent fields in the action: $F_{\mu\nu}$ and $\tilde{A}_\mu$\,, the latter being a Lagrange multiplier.
This parent action is invariant under
\begin{align}\label{eq:Maxwell_gauge}
    \delta\tilde{A}_\mu=\pd_\mu\tilde{\lambda}\,,
\end{align}
and its equations of motion are
\begin{subequations}
\begin{align}
    \mathcal{E}^{(\tilde{A})}_\mu&:=\tfrac{1}{2}\,\varepsilon_{\mu}{}^{\nu\rho\sigma}\pd_\nu F_{\rho\sigma}=0\,,\\
    \mathcal{E}^{(F)}_{\mu\nu}&:={-\tfrac12}F_{\mu\nu}+\tfrac12\,\varepsilon_{\mu\nu}{}^{\rho\sigma}\pd_\rho\tilde{A}_\sigma=0\,.
\end{align}
\end{subequations}
The first equation is solved by $F_{\mu\nu}=2\,\pd_{[\mu}A_{\nu]}$ up to $\delta A_\mu=\pd_\mu\lambda$\,, and substituting this into the parent action \eqref{eq:parent_Maxwell} leads to the original Maxwell action \eqref{eq:action_Maxwell}.
However, the second equation allows us to write $F_{\mu\nu}=\varepsilon_{\mu\nu}{}^{\rho\sigma}\pd_\rho\tilde{A}_\sigma$ and in this case the parent action reduces to a dual action for a dual field $\tilde{A}_\mu$ that takes the same form as \eqref{eq:action_Maxwell}.
The dual Maxwell action is
\begin{align}\label{eq:dual_Maxwell}
    S_\text{M}[\tilde{A}]=\int\!\rd^4x\,\Big({-\tfrac14}\tilde{F}[\tilde{A}]_{\mu\nu}\tilde{F}[\tilde{A}]^{\mu\nu}\Big)\,,
\end{align}
where $\tilde{F}[\tilde{A}]_{\mu\nu}:=2\,\pd_{[\mu}\tilde{A}_{\nu]}$ is the field strength of the dual field.
Under this duality, the electric and magnetic fields associated with $A_\mu$ and $\tilde{A}_\mu$ rotate into each other as usual:
\begin{align}
    &E_i:=-F_{0i}=-\tfrac12\,\varepsilon_{ijk}\tilde{F}^{jk}=:-\tilde{B}_i\,,&
    &B_i:=\tfrac12\,\varepsilon_{ijk}F^{jk}=-\tilde{F}_{0i}=:\tilde{E}_i\,,
\end{align}
where $i,j,\dots$ are spatial indices.

\subsection{Carrollian electromagnetism} \label{sec:Maxwell-Carroll}

We will now split the spatial and temporal components of the tensors in our relativistic parent action \eqref{eq:parent_Maxwell} so that it can be written in the form
\begin{align}\label{eq:parent_Maxwell_split}
    \hat{S}_{\text{M}}=\int\!\rd^4x\,\Big[{-\tfrac14}F_{ij}F^{ij}+\tfrac12F_{0i}F_0{}^i+\tfrac12\,\varepsilon^{0ijk}\Big(2F_{0i}\,\pd_j\tilde{A}_k+F_{ij}\big(\pd_0\tilde{A}_k-\pd_k\tilde{A}_0\big)\Big)\Big]\,.
\end{align}
We write the speed of light factors explicitly in each component as
\begin{align}
    \pd_0\;&\mapsto\frac1c\,\pd_t\,,& F_{0i}\;&\mapsto \frac1c\,F_{ti}\,,& \tilde{A}_0\;&\mapsto\frac1c\,\tilde{A}_t\,,& \varepsilon^{0ijk}\;&\mapsto\frac{1}{c}\,\varepsilon^{ijk} \,,
\end{align}
where $t$ is the (null) time-like direction and $i,j,k$ are spatial indices.
The parent action can now be written as
\begin{align}
    \hat{S}_{\text{M}}=c^2\!\int\!\rd t\,\rd^3x\,\Big[{-\tfrac14}F_{ij}F^{ij}+\tfrac1{2c^2}F_{ti}F_t{}^i+\tfrac1{2c^2}\,\varepsilon^{ijk}\Big(2F_{ti}\,\pd_j\tilde{A}_k+F_{ij}\big(\dot{\tilde{A}}_k-\pd_k\tilde{A}_t\big)\Big)\Big]\,,
\end{align}
where here and in the following, a dot denotes partial derivation with respect to $t$ and we have also introduced a compensating factor of $c^2$ in front of the action.
The gauge transformation \eqref{eq:Maxwell_gauge} now takes the form
\begin{align}\label{eq:Maxwell_gauge_split}
    \delta\tilde{A}_t&=\dot{\tilde{\lambda}}\,,&
    \delta\tilde{A}_i&=\pd_i\tilde{\lambda}\,.
\end{align}
Taking the $c\rightarrow0$ limit leads to the Carrollian parent action
\begin{align}\label{eq:parent_Maxwell_Carroll}
    \hat{S}_{\text{M}}^{(c\,\rightarrow\,0)}[F_{ti},F_{ij},\tilde{A}_t,\tilde{A}_i]=\int\!\rd t\,\rd^3x\,\Big[\tfrac12 F_{ti}F_t{}^i+\tfrac12\,\varepsilon^{ijk}\Big(2F_{ti}\,\pd_j\tilde{A}_k+F_{ij}\big(\dot{\tilde{A}}_k-\pd_k\tilde{A}_t\big)\Big)\Big]\,,
\end{align}
where, in addition to $\tilde A_t$ and $\tilde A_i$\,, $F_{ij}$ is now also a Lagrange multiplier. The equations of motion for $\tilde{A}_t$\,, $\tilde{A}_i$\,, $F_{ti}$\,, and $F_{ij}$ are given by
\begin{subequations}
\begin{align}
    \mathcal{E}^{(\tilde{A})}&:=\tfrac12\,\varepsilon^{ijk}\pd_iF_{jk}=0\,,\\
    \mathcal{E}^{(\tilde{A})}_i&:=\varepsilon_{i}{}^{jk}\left(\pd_jF_{tk}-\tfrac12\dot{F}_{jk}\right)=0\,,\\
    \mathcal{E}^{(F)}_i&:=F_{ti}+\varepsilon_{i}{}^{jk}\pd_j\tilde{A}_k=0\,,\\
    \mathcal{E}^{(F)}_{ij}&:=\tfrac12\,\varepsilon_{ij}{}^{k}\left(\dot{\tilde{A}}_k-\pd_k\tilde{A}_t\right)=0\,.
\end{align}
\end{subequations}

\paragraph{The electric theory.}
The first equation of motion is solved by $F_{ij} = F[A]_{ij} =2\,\pd_{[i}A_{j]}$ for some $A_i$ and consequently the second equation is solved by $F_{ti}= F[A]_{ti} = \dot{A}_i-\pd_iA_t$ for some $A_t$\,, defined up to $\delta A_i = \pd_i \lambda$ and $\delta A_t = \dot \lambda$\,.
The last equation of motion is also imposed, although we do not need it here.
As a result, using the equations of motion for $\tilde{A}_t$ and $\tilde{A}_i$\,, \eqref{eq:parent_Maxwell_Carroll} reduces to
\begin{align}\label{eq:Maxwell_Carroll_E}
    S_{\text{M-Elec.}}^{(c\,\rightarrow\,0)}[A_t,A_i]=\frac12\int\!\rd t\,\rd^3x\,\Big(\dot{A}_i-\pd_iA_t\Big)\Big(\dot{A}^i-\pd^iA_t\Big)\,.
\end{align}
This is the electric Carrollian limit of the Maxwell action, and its Hamiltonian form is given in equation (5.18) of \cite{Henneaux:2021yzg} which we reproduce here for convenience:
\begin{align}
    S_{\text{M-Elec.}}^{(c\,\rightarrow\,0)}[A_t,A_i,\pi_i]=\int\!\rd t\,\rd^3x\,\Big(\pi^i\dot{A}_i-\tfrac12\pi^i\pi_i+A_t\,\pd_i\pi^i\Big)\,.
\end{align}
The equations of motion of this theory in Maxwell form are
\begin{align}
    \nabla \cdot B &= 0\,,&
    \nabla \times E &= - \dot{B}\,,&
    \dot{E} &= 0\,,&
    &\nabla \cdot E = 0\,,
\end{align}
where $E_i := - F[A]_{ti}$ and $B_i = \frac12 \varepsilon_{ijk} F[A]^{jk} = \varepsilon_{ijk} \pd^j A^k$\,.
The first two equations follow from Bianchi identities and the last two follow varying \eqref{eq:Maxwell_Carroll_E} with respect to the fields $A_i$ and $A_t$ respectively.

\paragraph{The magnetic theory.}
Conversely, we can solve $F_{ti}$ through its own equation of motion and substitute $F_{ti}=-\varepsilon_{i}{}^{jk}\pd_j\tilde{A}_k$ inside the Carrollian parent action \eqref{eq:parent_Maxwell_Carroll} to obtain directly the magnetic theory
\begin{align}\label{eq:Maxwell_Carroll_M}
    S_{\text{M-Mag.}}^{(c\,\rightarrow\,0)}[\tilde{A}_t,\tilde{A}_i,\pi_i]=\int\!\rd t\,\rd^3x\,\Big(\pi^i\dot{\tilde{A}}_i-\tfrac14\tilde{F}[\tilde{A}]_{ij}\tilde{F}[\tilde{A}]^{ij}+\tilde{A}_t\,\pd_i\pi^i\Big)\,,
\end{align}
in its first-order form, where $\pi^i:=- \tfrac12\,\varepsilon^{ijk}F_{jk}$ and $\tilde{F}[\tilde{A}]_{ij}:=2\,\pd_{[i}\tilde{A}_{j]}$\,.
This magnetic action can be found in equation (5.13) of \cite{Henneaux:2021yzg}.
Its equations of motion in Maxwell form are
\begin{align}
    \nabla \cdot \tilde{B} &= 0\,,&
    \nabla \cdot \tilde{E} &= 0\,,&
    \nabla \times \tilde{B} &= \dot{\tilde{E}}\,,&
    \dot{\tilde{B}} &= 0\,,
\end{align}
where we defined $\tilde{E}_i := \pi_i$ and $\tilde{B}_i := \frac12 \varepsilon_{ijk} \tilde F[\tilde A]^{jk}$\,. The first is a Bianchi identity, the second and third follow from a variation with respect to $\tilde A_t$ and $\tilde A_i$ respectively, while the last is a consequence of the $\pi_i$ field equation, i.e.~$\dot{\tilde A}_i - \pd_i \tilde A_t = 0$\,.

We conclude that the electric and magnetic Carrollian limits of the Maxwell action \eqref{eq:action_Maxwell} are dual to each other in the usual sense that they arise from the same parent action by varying it with respect to different fields.
Moreover, the equations of motion are the same under the exchange $\{E,B\} \leftrightarrow \{-\tilde B, \tilde E\}$\,.
This is a direct consequence of the duality symmetry of the relativistic Maxwell theory in four dimensions.
Note that the electric and magnetic limits are both still gauge-invariant under \eqref{eq:Maxwell_gauge_split}.

\subsection{Galilean electromagnetism} \label{sec:Maxwell-Galilei}

The relativistic parent action \eqref{eq:parent_Maxwell_split} can be rewritten by including factors of $c$ explicitly:
\begin{subequations}
\begin{align}\label{eq:rescale_Maxwell_Galilei}
    \pd_0\;&\mapsto\frac1c\,\pd_t\,,&
    F_{0i}\;&\mapsto\frac1{c^2}\,F_{ti}\,,&
    F_{ij}\;&\mapsto\frac1c\,F_{ij}\,, \\
    \varepsilon^{0ijk}\;&\mapsto\frac1c\,\varepsilon^{ijk}\,,&
    \tilde{A}_0\;&\mapsto \tilde{A}_t\,, &\tilde{A}_i\;&\mapsto c\,\tilde{A}_i\,.
\end{align}
\end{subequations}
Substituting this into \eqref{eq:parent_Maxwell_split}, we obtain
\begin{align}
    \hat{S}_{\text{M}}=c^2\!\int\!\rd t\,\rd^3x\,\Big[{-\tfrac1{4c^2}}F_{ij}F^{ij}+ \tfrac1{2c^4}F_{ti}F_t{}^i+\tfrac1{2c^2}\,\varepsilon^{ijk}\Big(2F_{ti}\,\pd_j\tilde{A}_k+F_{ij}\big(\dot{\tilde{A}}_k-\pd_k\tilde{A}_t\big)\Big)\Big]\,,
\end{align}
where we have once again included a factor of $c^2$ in front so as to make the limit finite.
Taking the $c\to\infty$ limit leads to the Galilean parent action
\begin{align}\label{eq:parent_Maxwell_Galilei}
    \hat{S}_{\text{M}}^{(c\,\to\,\infty)}[F_{ti},F_{ij},\tilde{A}_t,\tilde{A}_i]=\int\!\rd t\,\rd^3x\,\Big[{-\tfrac14}F_{ij}F^{ij}+\tfrac12\,\varepsilon^{ijk}\Big(2F_{ti}\,\pd_j\tilde{A}_k+F_{ij}\big(\dot{\tilde{A}}_k-\pd_k\tilde{A}_t\big)\Big)\Big]\,.
\end{align}
This time we interpret $F_{ti}$ as a Lagrange multiplier in addition to $\tilde A_t$ and $\tilde A_i$\,.
The equations of motion for $\tilde{A}_t$\,, $\tilde{A}_i$\,, $F_{ti}$\,, and $F_{ij}$ are given by
\begin{subequations}
\begin{align}
    \mathcal{E}^{(\tilde{A})}&:=\tfrac12\,\varepsilon^{ijk}\pd_iF_{jk}=0\,,\\
    \mathcal{E}^{(\tilde{A})}_i&:=\varepsilon_{i}{}^{jk}\left(\pd_jF_{tk}-\tfrac12\dot{F}_{jk}\right)=0\,,\\
    \mathcal{E}^{(F)}_i&:=\varepsilon_{i}{}^{jk}\pd_j\tilde{A}_k=0\,,\\
    \mathcal{E}^{(F)}_{ij}&:=-\tfrac12F_{ij}+\tfrac12\,\varepsilon_{ij}{}^{k}\left(\dot{\tilde{A}}_k-\pd_k\tilde{A}_t\right)=0\,.
\end{align}
\end{subequations}
The parent action and its equations of motion obey the gauge transformations in \eqref{eq:Maxwell_gauge_split}.

\paragraph{The magnetic theory.}
As in the Carrollian case, the first two field equations are solved by $F_{ij} = F[A]_{ij} =2\,\pd_{[i}A_{j]}$ and $F_{ti} = F[A]_{ti} =\dot{A}_i-\pd_iA_t$ for some $A_i$ and $A_t$\,, while the third one is solved by $\tilde A_k = \pd_k \tilde \mu$ for some $\tilde \mu$ that can be gauged away using $\tilde \lambda$\,.
The parent action \eqref{eq:parent_Maxwell_Galilei} reduces to
\begin{align}\label{eq:Maxwell_Galilei_M}
    S_{\text{M-Mag.}}^{(c\,\rightarrow\,\infty)}[A_i]=\int\!\rd t\,\rd^3x\;\Big({-\tfrac14}F[A]_{ij}F[A]^{ij}\Big)\,.
\end{align}
The equation of motion for \eqref{eq:Maxwell_Galilei_M} is $\pd^iF[A]_{ij}=0$\,, therefore this action reproduces three of the four Maxwell equations for Galilean electromagnetism
\begin{align}
    \nabla \cdot B &= 0\,,&
    \nabla \times E &= - \dot{B}\,,&
    \nabla \times B &= 0\,,
\end{align}
where we noted the electric and magnetic fields $E_i:=-F[A]_{ti}$ and $B_i:=\tfrac12\,\varepsilon_{ijk}F[A]^{jk}$ (see \cite{LeBellac:1973unm}).
We do not recover $\nabla\cdot E=0$ in accordance with \cite{Bergshoeff:2022qkx}, where it is explained that constraints are lost in the Galilean limit of Maxwell theory -- see also \cite{Bergshoeff:2021bmc} for a more general discussion.

\paragraph{The electric theory.}
We can solve $F_{ij}$ through its own equation of motion and substitute $F_{ij}=\varepsilon_{ij}{}^k\left(\dot{\tilde{A}}_k-\pd_k\tilde{A}_t\right)$ inside the Galilean parent action \eqref{eq:parent_Maxwell_Galilei}, leading to the electric theory
\begin{align}\label{eq:Maxwell_Galilei_E}
    S_{\text{M-Elec.}}^{(c\,\rightarrow\,\infty)}[\tilde{A}_t,\tilde{A}_i,\tilde{B}_i]=\int\!\rd t\,\rd^3x\,\bigg[\tfrac12\Big(\dot{\tilde{A}}_i-\pd_i\tilde{A}_t\Big)\Big(\dot{\tilde{A}}^i-\pd^i\tilde{A}_t\Big)+\varepsilon^{ijk}\tilde{B}_i\pd_j\tilde{A}_k\bigg]\,.
\end{align}
Labelling $\tilde{B}_i := F_{ti}$\,, we find the equations of motion by varying \eqref{eq:Maxwell_Galilei_E} with respect to $\tilde{A}_t$\,, $\tilde{A}_i$\,, and $\tilde{B}_i$ to obtain, respectively,
\begin{subequations}
\begin{align}
    \mathcal{E}^{(\tilde A)} &:= \pd_i (\dot{\tilde{A}}^i - \pd^i \tilde{A}_t) = 0 \,,\\
    \mathcal{E}^{(\tilde A)}_i &:= -\left(\ddot{\tilde{A}}_i - \pd_i \dot{\tilde{A}}_t\right) - \varepsilon_i{}^{jk} \pd_j \tilde{B}_k= 0 \,,\\
    \mathcal{E}^{(B)}_i &:= \varepsilon_i{}^{jk} \pd_j \tilde{A}_k = 0 \,.
\end{align}
\end{subequations}
These equations translate, respectively, to
\begin{align}
    \nabla \cdot \tilde{E} &= 0 \,,&
    \nabla \times\tilde{B} &= \dot{\tilde{E}} \,,&
    \nabla \times \tilde{E} &= 0 \,,
\end{align}
where we noted $\tilde E_i := -\tilde F[\tilde A]_{ti} = -\dot{\tilde A}_i + \pd_i \tilde A_t$\,.
The last equation follows from $\nabla \times \tilde{A} = 0$\,.

We do not recover the last Maxwell equation $\nabla\cdot\tilde{B}=0$ for the same reason that we gave before, but we see that electric and magnetic theories are mapped into each other under the exchange $\{E,B\} \leftrightarrow \{-\tilde B, \tilde E\}$\,.

\section{Maxwell $p$-form electrodynamics} \label{sec:p-form}

We continue our analysis of non-Lorentzian duality with the more general case of the Maxwell $p$-form field in $D$ space-time dimensions, where $0 \leqslant p < D$\,.
The relativistic $p$-form action reads
\begin{align}\label{eq:action_Maxwell_p}
    S_{\text{M$(p)$}}[A]=\int\!\rd^Dx\,\Big({-\tfrac1{2(p+1)!}}F[A]_{\mu[p+1]}F[A]^{\mu[p+1]}\Big)\,,
\end{align}
where $F[A]_{[p+1]}:=\rd A_{[p]}$ is invariant under $\delta A_{[p]}=\rd\lambda_{[p-1]}$ when $p>0$\,.
Square brackets $\mu[n]$ denote $n$ antisymmetrised indices with unit weight, repeated indices are implicitly taken to be antisymmetrised, and differential forms\footnote{Our convention for a $p$-form $A$ is to write it as $A_{[p]}=\tfrac{1}{p!}A_{\mu_1\dots\mu_p}\rd x^{\mu_1}\wedge\dots\wedge\rd x^{\mu_p}$\,.} may be written in form notation with a subscript denoting their form degree.
It is straightforward to construct the parent action
\begin{align}\label{eq:parent_Maxwell_p}
\begin{split}
    \hat{S}_{\text{M$(p)$}}[F,\tilde{A}]=\int\!\rd^Dx\,\Big({-\tfrac1{2(p+1)!}}F_{\mu[p+1]}F^{\mu[p+1]}\hspace{60mm}\\
    +\tfrac1{(p+1)!(D-p-2)!}\,\varepsilon^{\mu[p+1]\nu[D-p-1]}F_{\mu[p+1]}\pd_\nu\tilde{A}_{\nu[D-p-2]}\Big)\,.
\end{split}
\end{align}
This action is invariant under $\delta \tilde{A}_{[D-p-2]}=\rd\tilde{\lambda}_{[D-p-3]}$ when $D-p-2>0$\,, and its equations of motion are given by
\begin{subequations}
\begin{align}
    \mathcal{E}^{(\tilde{A})}_{\mu[D-p-2]}&:=\varepsilon_{\mu[D-p-2]}{}^{\nu[p+2]}\pd_\nu F_{\nu[p+1]}=0\,,\\
    \mathcal{E}^{(F)}_{\mu[p+1]}&:={-\tfrac1{(p+1)!}}F_{\mu[p+1]}+\tfrac1{(p+1)!(D-p-2)!}\,\varepsilon_{\mu[p+1]}{}^{\nu[D-p-1]}\pd_\nu\tilde{A}_{\nu[D-p-2]}=0\,.
\end{align}
\end{subequations}
Just as in four dimensions, $\mathcal{E}^{(\tilde{A})}=0$ implies that $F_{[p+1]}=\rd A_{[p]}$ and so the parent action \eqref{eq:parent_Maxwell_p} reduces to the free relativistic $p$-form action \eqref{eq:action_Maxwell_p}.
The second equation $\mathcal{E}^{(F)}=0$ implies that $F_{[p+1]}=\star\,\rd\tilde{A}_{[D-p-2]}$ which is a gauge-invariant quantity.
Substituting this into the parent action leads to the free relativistic action $S_{\text{M}(D-p-2)}$ for the $(D-p-2)$-form $\tilde{A}_{[D-p-2]}$\,, taking the same form as \eqref{eq:action_Maxwell_p} with $p$ replaced by $(D-p-2)$\,.
Thus the $p$-form and the $(D-p-2)$-form theories are dual to each other in $D$ dimensions.

\subsection{Carrollian $p$-form electrodynamics} \label{sec:p-form-Carroll}

Splitting spatial and temporal components of the parent action \eqref{eq:parent_Maxwell_p}, it can be written as
\begin{align}
\begin{split}
    \hat{S}_{\text{M$(p)$}}=\int\!\rd^Dx\,\Big[{-\tfrac1{2(p+1)!}}F_{i[p+1]}F^{i[p+1]}-{\tfrac1{2p!}}F_{0i[p]}F^{0i[p]}\hspace{40mm}\\
    {}+\tfrac1{(p+1)!(D-p-2)!}\,\varepsilon^{0i[D-1]}\Big((p+1)F_{0i[p]}\pd_i\tilde{A}_{i[D-p-2]}\hspace{15mm}\\
    {}+(-1)^{p+1}F_{i[p+1]}\big(\pd_0\tilde{A}_{i[D-p-2]}-(D-p-2)\pd_i\tilde{A}_{0i[D-p-3]}\big)\Big)\Big]\,.\hspace{-9mm}
\end{split}
\end{align}
Factors of $c$ are made explicit using
\begin{align}
    \pd_0\;&\mapsto\frac1c\,\pd_t\,,& F_{0i[p]}\;&\mapsto \frac1c\,F_{ti[p]}\,,& \tilde{A}_{0i[D-p-3]}\;&\mapsto\frac1c\,\tilde{A}_{ti[D-p-3]}\,,& \varepsilon^{0i[D-1]}\;&\mapsto\frac{1}{c}\,\varepsilon^{i[D-1]}\,,
\end{align}
and taking the $c\rightarrow0$ limit leads to the Carrollian parent action
\begin{align}\label{eq:parent_Maxwell_Carroll_p}
\begin{split}
    \hat{S}^{(c\,\rightarrow\,0)}_{\text{M$(p)$}}=\int\!\rd t\,\rd^{D-1}x\,\Big[{-\tfrac1{2p!}}F_{ti[p]}F^{ti[p]}
    +\tfrac1{(p+1)!(D-p-2)!}\,\varepsilon^{i[D-1]}\Big((p+1)F_{ti[p]}\pd_i\tilde{A}_{i[D-p-2]}\\
    {}+(-1)^{p+1}F_{i[p+1]}\big(\dot{\tilde{A}}_{i[D-p-2]}-(D-p-2)\pd_i\tilde{A}_{ti[D-p-3]}\big)\Big)\Big]\,,
\end{split}
\end{align}
where $F_{i[p+1]}$ becomes a Lagrange multiplier.
The equations of motion for $\tilde{A}_{ti[D-p-3]}$\,, $\tilde{A}_{i[D-p-2]}$\,, $F_{ti[p]}$\,, and $F_{i[p+1]}$ are given by
\begin{subequations}
\begin{align}
    \mathcal{E}^{(\tilde{A})}_{i[D-p-3]}&:=\varepsilon_{i[D-p-3]}{}^{j[p+2]}\pd_jF_{j[p+1]}=0\,,\label{eq:eom1_Maxwell_Carroll_p}\\
    \mathcal{E}^{(\tilde{A})}_{i[D-p-2]}&:=\varepsilon_{i[D-p-2]}{}^{j[p+1]}\left((p+1)\pd_jF_{tj[p]}-\dot{F}_{j[p+1]}\right)=0\,,\label{eq:eom2_Maxwell_Carroll_p}\\
    \mathcal{E}^{(F)}_{i[p]}&:=F_{ti[p]}+\tfrac1{(D-p-2)!}\,\varepsilon_{i[p]}{}^{j[D-p-1]}\pd_j\tilde{A}_{j[D-p-2]}=0\,,\label{eq:eom3_Maxwell_Carroll_p}\\
    \mathcal{E}^{(F)}_{i[p+1]}&:=\varepsilon_{i[p+1]}{}^{j[D-p-2]}\left(\dot{\tilde{A}}_{j[D-p-2]}-(D-p-2)\pd_j\tilde{A}_{tj[D-p-3]}\right)=0\,,\label{eq:eom4_Maxwell_Carroll_p}
\end{align}
\end{subequations}
up to inessential factors.

\paragraph{The electric theory.}
As before, the first equation \eqref{eq:eom1_Maxwell_Carroll_p} implies that $F_{i[p+1]}=(p+1)\,\pd_iA_{i[p]}$ for some $A_{i[p]}$ up to $\delta A_{i[p]}=p\,\pd_i\lambda_{i[p-1]}$ when $p>0$\,.
The second equation \eqref{eq:eom2_Maxwell_Carroll_p} is then solved by $F_{ti[p]}=\dot{A}_{i[p]}-p\,\pd_iA_{ti[p-1]}$ for some $A_{ti[p-1]}$ up to $\delta A_{ti[p-1]}=\dot{\lambda}_{i[p-1]}-(p-1)\pd_i\lambda_{i[p-2]t}$\,, and the last one is imposed as well, although it does not lead to any particularly interesting constraints or dynamics. The parent action \eqref{eq:parent_Maxwell_Carroll_p} reduces to the action for the electric Carrollian $p$-form:
\begin{align}\label{eq:Maxwell_Carroll_p_E}
    S^{(c\,\rightarrow\,0)}_{\text{M$(p)$-Elec.}}=\frac1{2p!}\int\!\rd t\,\rd^{D-1}x\,\Big(\dot{A}_{i[p]}-p\,\pd_iA_{ti[p-1]}\Big)\Big(\dot{A}^{i[p]}-p\,\pd^iA_t{}^{i[p-1]}\Big)\,.
\end{align}
The equivalent first-order form of $S^{(c\,\rightarrow\,0)}_{\text{M$(p)$-Elec.}}$ found in \cite{Henneaux:2021yzg} is
\begin{align}
    S^{(c\,\rightarrow\,0)}_{\text{M$(p)$-Elec.}}=\int\!\rd t\,\rd^{D-1}x\,\Big(\pi^{i[p]}\dot{A}_{i[p]}-\tfrac{p!}{2}\pi_{i[p]}\pi^{i[p]}+p\,\pd_i\pi^{i[p]}A_{ti[p-1]}\Big)\,.
\end{align}

\paragraph{The magnetic theory.}
Using the third equation of motion \eqref{eq:eom3_Maxwell_Carroll_p} we find that \eqref{eq:parent_Maxwell_Carroll_p} reduces to the action for the magnetic Carrollian $(D-p-2)$-form that can be found in \cite{Henneaux:2021yzg}\,:
\begin{align}\label{eq:Maxwell_Carroll_p_M}
\begin{split}
    S^{(c\,\rightarrow\,0)}_{\text{M$(D\!-\!p\!-\!2)$-Mag.}}
    =\int\!\rd t\,\rd^{D-1}x\,\Big(\pi^{i[D-p-2]}\dot{\tilde{A}}_{i[D-p-2]}-\tfrac1{2(D-p-1)!}\tilde{F}[\tilde{A}]_{i[D-p-1]}\tilde{F}[\tilde{A}]^{i[D-p-1]}\\
    {}+(D-p-2)\,\pd_i\pi^{i[D-p-2]}\tilde{A}_{ti[D-p-3]}\Big)\,,
\end{split}
\end{align}
where $\tilde{F}[\tilde{A}]_{i[p+1]}:=(p+1)\pd_i\tilde{A}_{i[p]}$ and $\pi_{i[D-p-2]}:=\tfrac{(-1)^{p+1}}{(p+1)!(D-p-2)!}\,\varepsilon^{j[p+1]}{}_{i[D-p-2]}F_{j[p+1]}$\,.
We have shown that the electric Carrollian $p$-form is dual to the magnetic Carrollian $(D-p-2)$-form in $D$ dimensions.

\subsection{Galilean $p$-form electrodynamics} \label{sec:p-form-Galilei}

The relativistic parent action for the Maxwell $p$-form \eqref{eq:parent_Maxwell_p} with split spatial and temporal components can be rewritten by making factors of $c$ explicit:
\begin{subequations}
\begin{align}
    \pd_0\;&\mapsto\frac1c\,\pd_t\,,&
    F_{0i[p]}\;&\mapsto\frac1{c^2}\,F_{ti[p]}\,,&
    F_{i[p+1]}\;&\mapsto\frac1c\,F_{i[p+1]}\,,\\
    \varepsilon^{0i[D-1]}\;&\mapsto\frac1c\,\varepsilon^{i[D-1]}\,,&\tilde{A}_{0i[D-p-3]}\;&\mapsto \tilde{A}_{ti[D-p-3]}\,,&\tilde{A}_{i[D-p-2]}\;&\mapsto c\,\tilde{A}_{i[D-p-2]}\,.
\end{align}
\end{subequations}
Taking the $c\rightarrow\infty$ limit leads to the Galilean parent action
\begin{align}\label{eq:parent_Maxwell_Galilei_p}
    \hat{S}^{(c\,\rightarrow\,\infty)}_{\text{M$(p)$}}=\int\!\rd t\,\rd^{D-1}x\,\Big[{-\tfrac1{2(p+1)!}}F_{i[p+1]}F^{i[p+1]}
    +\tfrac1{(p+1)!(D-p-2)!}\,\varepsilon^{i[D-1]}\Big((p+1)F_{ti[p]}\pd_i\tilde{A}_{i[D-p-2]}\hspace{-10mm}\nonumber\\
    {}+(-1)^{p+1}F_{i[p+1]}\big(\dot{\tilde{A}}_{i[D-p-2]}-(D-p-2)\pd_i\tilde{A}_{ti[D-p-3]}\big)\Big)\Big]\,,
\end{align}
where $F_{ti[p]}$ becomes a Lagrange multiplier. The equations of motion for $\tilde{A}_{ti[D-p-3]}$\,, $\tilde{A}_{i[D-p-2]}$\,, and $F_{ti[p]}$ are given by
\begin{subequations}
\begin{align}
    \mathcal{E}^{(\tilde{A})}_{i[D-p-3]}&:=\varepsilon_{i[D-p-3]}{}^{j[p+2]}\pd_jF_{j[p+1]}=0\,,\label{eq:eom1_Maxwell_Galilei_p}\\
    \mathcal{E}^{(\tilde{A})}_{i[D-p-2]}&:=\varepsilon_{i[D-p-2]}{}^{j[p+1]}\left((p+1)\pd_jF_{tj[p]}-\dot{F}_{j[p+1]}\right)=0\,,\label{eq:eom2_Maxwell_Galilei_p}\\
    \mathcal{E}^{(F)}_{i[p]}&:=\varepsilon_{i[p]}{}^{j[D-p-1]}\pd_j\tilde{A}_{j[D-p-2]}=0\,,\label{eq:eom3_Maxwell_Galilei_p}
\end{align}
and the one for $F_{i[p+1]}$ by
\begin{align}
    \hspace{-3mm}\mathcal{E}^{(F)}_{i[p+1]}&:=-F_{i[p+1]}+\tfrac{(-1)^p}{(D-p-2)!}\varepsilon_{i[p+1]}{}^{j[D-p-2]}\left(\dot{\tilde{A}}_{j[D-p-2]}-(D-p-2)\pd_j\tilde{A}_{tj[D-p-3]}\right)=0\,,\label{eq:eom4_Maxwell_Galilei_p}
\end{align}
\end{subequations}
all up to inessential factors.

\paragraph{The magnetic theory.}

The first two equations \eqref{eq:eom1_Maxwell_Galilei_p} and \eqref{eq:eom2_Maxwell_Galilei_p} are once again solved by $F_{i[p+1]}=(p+1)\,\pd_iA_{i[p]}$ and $F_{ti[p]}=\dot{A}_{i[p]}-p\,\pd_iA_{ti[p-1]}$ up to gauge transformations
\begin{align}
    \delta A_{i[p]}&=p\,\pd_i\lambda_{i[p-1]}\,,&
    \delta A_{ti[p-1]}&=\dot{\lambda}_{i[p-1]}-(p-1)\pd_i\lambda_{i[p-2]t}\,.
\end{align}
The third is solved by $\tilde A_{j[D-p-2]} = (D-p-2) \pd_j \tilde\mu_{j[D-p-3]}$, as long as $D \geqslant p + 3$\,, that can be gauged away using the transformation of $\tilde A_{j[D-p-3]}$\,. The parent action \eqref{eq:parent_Maxwell_Galilei_p} now reduces to
\begin{align}\label{eq:Maxwell_Galilei_p_M}
    S^{(c\,\rightarrow\,\infty)}_{\text{M$(p)$-Mag.}}=\int\!\rd t\,\rd^{D-1}x\,\Big[{-\tfrac1{2(p+1)!}}F[A]_{i[p+1]}F[A]^{i[p+1]}\Big]\,,
\end{align}
where $F[A]_{i[p+1]}:=(p+1)\pd_iA_{i[p]}$\,.
This matches the magnetic Galilean $p$-form action of \cite{Bergshoeff:2022qkx}.

\paragraph{The electric theory.}

Using equation \eqref{eq:eom4_Maxwell_Galilei_p} we find that the parent action \eqref{eq:parent_Maxwell_Galilei_p} reduces to the action for the electric Galilean $(D-p-2)$-form that can be found in \cite{Henneaux:2021yzg}\,:
\begin{align}\label{eq:Maxwell_Galilei_p_E}
\begin{split}
    S_{\text{M$(p)$-Elec.}}^{(c\,\rightarrow\,\infty)}=\int\!\rd t\,\rd^{D-1} x\,\Big[\tfrac1{2(D-p-2)!}\Big(\dot{\tilde{A}}_{i[D-p-2]}-(D-p-2)\pd_i\tilde{A}_{ti[D-p-3]}\Big)^2\hspace{10mm}\\
    {}+\chi^{i[D-p-1]}\pd_i\tilde{A}_{i[D-p-2]}\Big]\,,
\end{split}
\end{align}
where $\chi^{i[D-p-1]}:=\tfrac{1}{(p+1)!(D-p-2)!}\,\varepsilon^{j[p]i[D-p-1]}F_{tj[p]}$\,.
Thus the electric Galilean $p$-form is dual to the magnetic Galilean $(D-p-2)$-form in $D$ dimensions.

\subsection{Example\,: scalar field in two dimensions} \label{sec:scalar}

The case of the conformally coupled Carrollian scalar field in $d+1$ dimensions over $\mathbb R \times S^d$ was studied in detail in \cite{Bekaert:2024itn}.
It was found that the two theories, although different in appearance, admit the same space of solutions and even the same equations of motion up to a non-local transformation.
Here, we focus on the case of the two-dimensional scalar field which is dual to itself in the relativistic context, and we show that electric and magnetic non-Lorentzian theories are indeed dual to each other off-shell.

The relativistic scalar field in $1+1$ dimensions is described by the Klein-Gordon action
\begin{align}\label{eq:action_scalar}
    S_\text{KG}[\phi] = \int\!\rd^2 x \Big({-\tfrac12}\,\pd_\mu\phi\, \pd^\mu \phi\Big) \,.
\end{align}
As always, we will neglect boundary terms.
The parent action is given by
\begin{align}
    \hat S_\text{KG}[F,\tilde\phi] = \int\!\rd^2 x \left( {-\tfrac12}F^\mu F_\mu + \varepsilon^{\mu\nu}F_\mu\pd_\nu\tilde\phi \right),
\end{align}
and its equations of motion are
\begin{equation}
    \mathcal E^{(\tilde\phi)} := \varepsilon^{\mu\nu} \pd_\mu F_\nu = 0 \,,\quad 
    \mathcal E^{(F)}_\mu := F_\mu-\varepsilon_\mu{}^\nu \pd_\nu\tilde\phi = 0 \,.
\end{equation}
The first equation is solved by $F_\mu=\pd_\mu\phi$ and substituting this back leads directly to \eqref{eq:action_scalar}.
The second yields, up to a boundary term, an identical dual action.
\begin{align}
    S_\text{KG}[\tilde{\phi}] = \int\!\rd^2x \,\Big({-\tfrac12}\pd_\mu\tilde\phi\,\pd^\mu\tilde\phi\Big)\,.
\end{align}

\paragraph{Carrollian contraction.}

Splitting the tensors into time and space components, then rescaling them appropriately and taking the Carrollian limit, the parent action becomes
\begin{align}
    \hat{S}^{(c\,\rightarrow\,0)}_\text{KG}[F,\tilde\phi] = &\int\!\rd t\,\rd x \left(\tfrac12 F_t{}^2 + F_t\,\pd_x\tilde{\phi} - F_x \dot{\tilde{\phi}}\right) .
\end{align}
The equations of motion of this Carrollian parent action are given by
\begin{subequations}
\begin{align}
    \mathcal E^{(\tilde\phi)} &:= \dot{F}_x-\pd_xF_t = 0 \,,\\
    \mathcal E^{(F_x)} &:= - \dot{\tilde \phi} = 0 \,,\\
    \mathcal E^{(F_t)} &:= F_t+\pd_x\tilde\phi = 0 \,.
\end{align}
\end{subequations}
Solving the first two equations as $F_\mu=\pd_\mu\phi$ and $\tilde \phi = \tilde \phi(x)$ and substituting back yields the electric action of \cite{Henneaux:2021yzg}:
\begin{align}
    S^{(c\,\rightarrow\,0)}_\text{KG-Elec.}[\phi] = \int \rd t \, \rd x \left( \tfrac12 \dot{\phi}^2\right) .
\end{align}
Conversely, varying with respect to $F_t$ and imposing the resulting equation of motion yields $F_t=-\pd_x\tilde{\phi}$\,, leading to the magnetic action of \cite{Henneaux:2021yzg}:
\begin{align}
    S^{(c\,\rightarrow\,0)}_\text{KG-Mag.}[\tilde\phi,\pi] = \int\!\rd t\,\rd x \left(\pi\,\dot{\tilde\phi} - \tfrac12 \pd_x\tilde\phi \, \pd_x \tilde\phi\right) ,
\end{align}
where $\pi:=-F_x$ is the conjugate momentum.

One might be surprised that such a duality between electric and magnetic Carrollian scalars exists, since it was found in \cite{deBoer:2021jej} that the electric and magnetic theories have different ground states and Hilbert spaces.
However, the analysis of that reference was performed for a massive electric scalar, while our duality is only valid for the case of the massless scalar field.

\paragraph{Galilean contraction.}

We can use the Carroll-Galilei duality in two dimensions and display the result immediately by exchanging $t$ and $x$\,. The parent action reads
\begin{align}
    \hat{S}^{(c\,\rightarrow\,\infty)}_\text{KG}[F,\tilde\phi] = &\int\!\rd t\,\rd x \left(\tfrac12 F_x{}^2 + F_x\,\dot{\tilde{\phi}} - F_t \pd_x \tilde{\phi} \right) ,
\end{align}
and
varying with respect to the various fields yields either an `electric' action
\begin{align}
    S^{(c\,\rightarrow\,\infty)}_\text{KG-Elec.}[\phi] = \int \rd t \, \rd x \,\Big( \tfrac12 \pd_x \phi\,\pd_x \phi \Big)\,,
\end{align}
or a `magnetic' action
\begin{align}
    S^{(c\,\rightarrow\,\infty)}_\text{KG-Mag.}[\tilde\phi,\pi] = \int\!\rd t\,\rd x \left(\pi\,\pd_x \tilde\phi - \tfrac12 \dot{\tilde\phi}^2\right) .
\end{align}

\section{Linearised gravity in four dimensions} \label{sec:Fierz-Pauli}

We will now consider duality symmetries for linearised gravity in four dimensions.
This case is particularly interesting because the dual of the Fierz-Pauli field is a rank-two symmetric tensor called the dual graviton whose dynamics is again that of Fierz-Pauli \cite{Hull:2000zn,Hull:2001iu,Boulanger:2003vs,Curtright:1980yk,Henneaux:2004jw}.

Let us start by recalling the Fierz-Pauli action on Minkowski background in terms of the linearised graviton $h_{ab}$ (we set $16\pi G_N = 1$)
\begin{equation} \label{eq:FP}
    S_\text{FP}[h] = \int\!\rd^4x\,\Big({-\tfrac14}\pd_a h_{bc}\pd^a h^{bc}+\tfrac12\pd^a h_{ab}\pd_c h^{bc}-\tfrac12\pd^a h_{ab}\pd^b h+\tfrac14\pd_a h \pd^a h \Big)\,,
\end{equation}
where indices are raised and lowered with $\eta_{ab}$ and $h := h_{ab} \eta^{ab}$\,. This second-order action for linearised gravity in four dimensions can also be written in terms of a linearised vielbein $e_{a|b}$ as
\begin{align}\label{eq:Fierz-Pauli}
    S_{\text{FP}}[e]=\int\!\rd^4x\,\Big({-\tfrac14}\,\Omega[e]_{ab|c}\,\Omega[e]^{ab|c}-\tfrac12\,\Omega[e]_{ab|c}\,\Omega[e]^{ac|b}+\,\Omega[e]_{ab|}{}^b\,\Omega[e]^{ac|}{}_c\Big)\,,
\end{align}
where $\Omega[e]_{ab|}{}^c := 2\,\pd_{[a}e_{b]|}{}^c$\,.
The first indices of $e_{a|b}$ and $\Omega_{ab|c}$ are to be thought of as form indices (raised and lowered with the background vielbein $\delta_a^b$), while the last are frame indices.
This action is invariant under $\delta e_{a|b}=\pd_{a}\xi_{b}+\lambda_{ab}$\,, where the new parameter $\lambda_{ab} = \lambda_{[ab]}$ corresponds to (linearised) local Lorentz transformations.
One can go back to the original Fierz-Pauli action by imposing to the so-called `metric gauge', where one shifts away the antisymmetric part of $e_{a|b}$ and identifies\footnote{The factor of $2$ comes from $g_{\mu\nu} = \eta_{\mu\nu} + h_{\mu\nu} + \mathcal O(h^2)$ where $h_{\mu\nu} = \eta_{ab} (\delta_{\mu}^a e_{\nu|}{}^b + e_{\mu|}{}^a \delta_\nu^b) = 2\,e_{(\mu|\nu)}$\,.} $h_{ab} = 2\,e_{(a|b)}$\,.
One can construct the relativistic parent action
\begin{align}\label{eq:parent_Fierz-Pauli}
    \hat{S}_{\text{FP}}\big[\Omega,\tilde{e}\big]=\int\!\rd^4x\,\Big({-\tfrac14}\,\Omega_{ab|c}\,\Omega^{ab|c}-\tfrac12\,\Omega_{ab|c}\,\Omega^{ac|b}+\Omega_{ab|}{}^b\,\Omega^{ac|}{}_c+\varepsilon^{abde}\,\Omega_{ab|c}\,\pd_d\tilde{e}_{e|}{}^c\Big)\,,
\end{align}
where $\Omega_{ab|}{}^c=\Omega_{[ab]|}{}^c$ is now an independent field \cite{West:2001as,West:2002jj,Boulanger:2003vs,Boulanger:2022arw}.
The equations of motion are
\begin{subequations}
\begin{align}
    \mathcal{E}^{(\tilde{e})}_{a|b}&:=\tfrac12\,\varepsilon_a{}^{cde}\,\pd_c\Omega_{de|b}=0\,, \label{eq:variation-tilde-e}\\
    \mathcal{E}^{(\Omega)}_{ab|c}&:=-\tfrac12\,\Omega_{ab|c}+\Omega_{c[a|b]}-2\,\eta_{c[a}\Omega_{b]d|}{}^d+\varepsilon_{ab}{}^{de}\,\pd_d\tilde{e}_{e|c}=0\,.
\end{align}
\end{subequations}
The first equation is solved by $\Omega_{ab|c}=2\,\pd_{[a}e_{b]|c}$\,, leading back to \eqref{eq:Fierz-Pauli}.
The second equation is solved by
\begin{align}
    \Omega_{ab|c}=-2\,\varepsilon_{cde[a}\,\pd^d\tilde{e}^{e|}{}_{b]}+\eta_{c[a}\varepsilon_{b]def}\,\pd^d\tilde{e}^{e|f}\,.
\end{align}
Substituting this back into the parent action and using the algebraic gauge symmetry of $\tilde{e}_{a|b}$ to shift away its antisymmetric component, one recovers the Fierz-Pauli action once again but this time in terms of the dual graviton $\tilde{h}_{ab}:=2\,\tilde{e}_{(a|b)}$\,.

\paragraph{Gauge symmetries.}

The parent action $\hat S_\text{FP}$ is invariant under the gauge transformations
\begin{align}
    &\delta\Omega_{ab|c}= 2\,\pd_{[a}\lambda_{b]c}\,,&
    &\delta\tilde{e}_{a|b}=\pd_a\tilde{\xi}_b-\tfrac12\,\varepsilon_{abcd}\lambda^{cd}\,.
\end{align}
When varying with respect to $\tilde{e}_{a|b}$\,, one sees that $\Omega_{ab|c}$ becomes the field strength (similar to the spin connection) of a vielbein $e_{a|b}$ enjoying the gauge symmetry $\delta e_{a|b} = \pd_a \xi_b + \lambda_{ab}$\,.
If we eliminate $\Omega_{ab|c}$ instead, then we get a theory for the (dual) vielbein $\tilde{e}_{a|b}$ which enjoys the same gauge transformation but with the parameters $\tilde \xi_a$ and $\tilde \lambda_{ab}:=-\tfrac12\,\varepsilon_{abcd}\,\lambda^{cd}$\,.

\subsection{Carrollian linearised gravity} \label{sec:Fierz-Pauli-Carroll}

\paragraph{Carrollian parent action.}

Let us open the potential term (i.e.~terms quadratic in $\Omega_{ab|c}$) in the relativistic parent action \eqref{eq:parent_Fierz-Pauli} by splitting the tensors into space and time components:
\begin{align}\label{eq:parent_Fierz-Pauli_split}
\hat{S}_{\text{FP}}\big[\Omega,\tilde{e}\big]=\int\!\rd^4x\,&\Big[{-\tfrac14}\,\Omega_{ij|0}\,\Omega^{ij|0} + \Omega_{ij|0}\,\Omega^{0i|j} - 2\,\Omega_{i0|0}\,\Omega^{ij|}{}_j -\Omega_{0(i|j)}\,\Omega^{0(i|j)}+\Omega_{0i|}{}^i\,\Omega^{0j|}{}_j \nonumber\\
    &\hspace{10mm}{}+ \Omega_{ij|k}\left({-\tfrac14}\,\Omega^{ij|k}-{\tfrac12}\,\Omega^{ik|j} + \delta^{jk} \Omega^{il|}{}_l \right)\hspace{30mm}\\
    &\hspace{-25mm}{}+\varepsilon^{0ijk}\Big(2\,\Omega_{0i|l}\,\pd_j\tilde{e}_{k|}{}^l-2\,\Omega_{i0|0}\,\pd_j\tilde{e}_{k|}{}^0 -\Omega_{ij|0}\Big(\pd_0\tilde{e}_{k|}{}^0-\pd_k\tilde{e}_{0|}{}^0\Big) +\Omega_{ij|l}\Big(\pd_0\tilde{e}_{k|}{}^l-\pd_k\tilde{e}_{0|}{}^l\Big) \Big)\Big]\,. \nonumber
\end{align}
In order to take the Carrollian contraction we first write the factors of $c$ explicitly using
\begin{subequations}
\begin{align}
    \pd_0\;&\mapsto\frac1c\,\pd_t\,,&
    \tilde{e}_{0|}{}^i\;&\mapsto\frac1c\,\tilde{e}_{t|}{}^i\,,&
    \tilde{e}_{i|}{}^0\;&\mapsto c\,\tilde{e}_{i|}{}^t\,,&
    \tilde{e}_{0|}{}^0\;&\mapsto\tilde{e}_{t|}{}^t\,,\\
    \varepsilon^{0ijk}\;&\mapsto\frac1c\,\varepsilon^{ijk}\,,&
    \Omega_{ij|0}\;&\mapsto\frac1c\,\Omega_{ij|t}\,,&
    \Omega_{0i|j}\;&\mapsto\frac1c\,\Omega_{ti|j}\,,&
    \Omega_{i0|0}\;&\mapsto\frac1{c^2}\,\Omega_{it|t}\,.
\end{align}
\end{subequations}
Rescaling the action by $c^2$ and taking the $c \to 0$ limit, the Carrollian parent action reads
\begin{align}\label{eq:parent_FP_Carroll}
    \hat{S}_\text{FP}^{(c\,\to\,0)}\big[\Omega,\tilde{e}\big] = \int\!\rd t\,\rd^3x\,&\Big(\tfrac14\,\Omega_{ij|t}\,\Omega^{ij|}{}_t - \Omega_{ij|t}\,\Omega_t{}^{i|j} - 2\,\Omega_{it|t}\,\Omega^{ij|}{}_j+\Omega_{t(i|j)}\,\Omega_t{}^{(i|j)}-\Omega_{ti|}{}^i\,\Omega_{tj|}{}^j \\
    &\hspace{-15mm}{}+\varepsilon^{ijk}\Big(2\,\Omega_{it|t}\,\pd_j\tilde{e}_{k|t}+2\,\Omega_{ti|l}\,\pd_j\tilde{e}_{k|}{}^l-\Omega_{ij|t}\Big(\dot{\tilde{e}}_{k|t}-\pd_k\tilde{e}_{t|t}\Big)+\Omega_{ij|}{}^l\Big(\dot{\tilde{e}}_{k|l}-\pd_k\tilde{e}_{t|l}\Big)\Big)\,, \nonumber
\end{align}
where the purely spatial terms $\Omega_{ij|k}$ in the potential do not appear except as a trace.
One can check that this action is invariant under the gauge transformations
\begin{align}
    \delta\Omega_{ab|c} &= 2\,\pd_{[a}\lambda_{b]c}\,,&
    \delta_{\tilde{\xi}}\tilde{e}_{a|b} &= \pd_a \tilde\xi_b \,,& \delta_{\lambda}\tilde{e}_{i|t} &=- \tfrac12\,\varepsilon_i{}^{jk}\lambda_{jk}\,,&
    \delta_{\lambda}\tilde{e}_{i|j} &=-\varepsilon_{ij}{}^k \lambda_{kt} \,.
\end{align}
The transformation law of the auxiliary dual vielbein field $\tilde e_{a|b}$ under $\tilde \lambda_i = - \frac12 \varepsilon_{i}{}^{jk}\,\lambda_{jk}$ shows that it enjoys (linearised) Carrollian boost symmetry: $\delta_{\tilde \lambda} \tilde e_{i|t} = \tilde \lambda_i$\,.
In contrast, if we vary the Carrollian parent action with respect to $\tilde e_{a|b}$\,, then the `spin connection' $\Omega_{ab|c}$ can be expressed in terms of a new linearised frame $e_{a|b}$ enjoying the same symmetries as a relativistic linearised frame, i.e.~$\delta_\lambda e_{a|b} = \lambda_{ab}$\,.
One may be puzzled by the fact that a Carrollian field enjoys the same gauge symmetries as the ones of a relativistic theory.
This peculiarity of the linearised electric theory is in agreement with the Hamiltonian formulation of \cite{Henneaux:2021yzg}, where it can be observed that the metric and its conjugate momentum are invariant under rigid \emph{Poincar\'e} symmetries.

\paragraph{Recovering electric Carroll gravity.}

We vary the Carrollian parent action with respect to all Lagrange multipliers, that is $\tilde{e}_{a|b}$ and $\hat{\Omega}_{ij|k} = \left.\Omega_{ij|k}\right|_\text{trace-free}$ and solve the same equation as \eqref{eq:variation-tilde-e} to obtain $\Omega_{ab|c} = 2\,\pd_{[a} e_{b]|c}$ and some inessential constraint on $\tilde{e}_{a|b}$\,.
Substituting this back, imposing the an equivalent of the metric gauge $e_{[i|j]} = 0$ and $e_{i|t} = 0$ thanks to the parameters $\lambda_i$ and $\lambda_{ij}$\,, and identifying $h_{ij}:=2\,e_{(i|j)}$\,, $h_{tt}:=2\,e_{t|t}$\,, and $h_{ti}=h_{it}:=e_{t|i}$\,, we obtain an action that takes the form\footnote{The last identification follows from $h_{ab} = 2\,e_{(a|b)}$ with $e_{i|t} = 0$\,.}
\begin{align} \label{eq:FP-Carroll-electric}
    S^{(c\,\to\,0)}_\text{FP-Elec.}[h_{ij},h_{ti},h_{tt}] = \int\!\rd t\,\rd^3x\,\Big[K_{ij} K^{ij} - K^2 + \tfrac{1}{2}\,h_{tt}\, \pd_i \! \left(\pd^i h_j{}^j-\pd_j h^{ij}\right) \Big]\,,
\end{align}
where $K_{ij}=K_{(ij)}:=-\pd_{[t} e_{i]|j} - \pd_{[t} e_{j]|i} = - \frac12 \dot h_{ij} + \pd_{(i} h_{j)t}$ and $K = K^i{}_i$\,. The tensor $K_{ij}$ agrees with the linearisation of the extrinsic curvature and therefore this theory is nothing but the linearisation of the `electric' action of \cite{Henneaux:1979vn} (see also \cite{Henneaux:2021yzg,Hansen:2021fxi}).

Imposing the gauge $e_{i|t} = 0$ fixes the parameter of local boosts $\lambda_{ti} = \pd_i \xi_t$\,.
As a result, the metric transforms as \mbox{$\delta_\xi h_{ij} = 2\,\pd_{(i} \xi_{j)}$}\,, \mbox{$\delta_\xi h_{ti} = \dot \xi_i + \pd_i \xi_t$}\,, and \mbox{$\delta_\xi h_{tt} = 2\,\pd_t \xi_t$}  under linearised diffeomorphisms, i.e.~as a relativistic linearised metric.
Therefore, the rigid symmetries of this theory are Poincar\'e and not Carroll.

\paragraph{Recovering magnetic Carroll gravity.}

Now we vary the parent action \eqref{eq:parent_FP_Carroll} with respect to the independent field $\Omega_{ab|c}$\,, leaving out the part which is a Lagrange multiplier, and solve the equations of motion to express $\Omega_{ab|c}$ in terms of $\tilde{e}_{a|b}$\,.
Varying component by component in the parent action (save for the trace-free part of $\Omega_{ij|k}$) we find
\begin{subequations}
\begin{align}
    \frac{\delta \hat S}{\delta \Omega_{ij|t}} = 0 &\quad\Longrightarrow\quad  \tfrac12\,\Omega^{ij|}{}_t - \Omega_t{}^{[i|j]} - \varepsilon^{ijk} \left(\dot{\tilde e}_{k|t}-\pd_k\tilde{e}_{t|t}\right)=0 \,,\\
    \frac{\delta \hat S}{\delta \Omega_{t[i|j]}} = 0 &\quad\Longrightarrow\quad -\Omega^{ij|}{}_t + 2\,\varepsilon^{kl[i} \pd_k \tilde e_{l|}{}^{j]}=0 \,,\\
    \frac{\delta \hat S}{\delta \Omega_{t(i|j)}} = 0 &\quad\Longrightarrow\quad 2\,\Omega_t{}^{(i|j)} - 2\,\delta^{ij} \Omega_{tk|}{}^k + 2\,\varepsilon^{kl(i} \pd_k \tilde e_{l|}{}^{j)}=0 \,,\\
    \frac{\delta \hat S}{\delta \Omega_{it|t}} = 0 &\quad\Longrightarrow\quad -2\,\Omega^{ij|}{}_j + 2\,\varepsilon^{ijk} \pd_j \tilde e_{k|t}=0 \,,\\
    \frac{\delta \hat S}{\delta \Omega_{ij|}{}^j} = 0 &\quad\Longrightarrow\quad -2\,\Omega^i{}_{t|t} - \varepsilon^{ijk} \left(\dot{\tilde e}_{j|k}-\pd_j\tilde{e}_{t|k}\right) = 0 \,.
\end{align}
\end{subequations}
Observe that these variations form a closed set in the sense that varying the parent action with respect to this set of components allows one to fix them completely in terms of $\tilde{e}_{a|b}$\,.
We shall also fix a gauge wherein $\tilde e_{[i|j]}$ and $\tilde e_{i|t}$ are set to zero, and identify $\tilde{h}_{ij}:=2\,\tilde{e}_{(i|j)}$\,, $\tilde{h}_{it}=\tilde{h}_{ti}:=\tilde{e}_{t|i}$\,, and $\tilde{h}_{tt}:=2\,\tilde{e}_{t|t}$\,.
Solving the above equations of motion, we find
\begin{subequations}
\begin{align}
    \Omega_{ij|}{}^j&=0\,,& \Omega_{ij|t}&=\varepsilon_{kl[i}\,\pd^k\tilde{h}^l{}_{j]}\,,&
    \Omega_{t[i|j]}&=\tfrac12\,\varepsilon_{kl[i}\,\pd^k\tilde{h}^l{}_{j]}+\tfrac12\,\varepsilon_{ijk}\,\pd^k\tilde{h}_{tt}\,, \\
    \Omega_{ti|}{}^i&=0 \,,&
    \Omega_{it|t}&=\tfrac12\,\varepsilon_{ijk}\,\pd^j\tilde{h}^k{}_t\,,&
    \Omega_{t(i|j)}&=-\tfrac12\,\varepsilon_{kl(i}\,\pd^k\tilde{h}^l{}_{j)}\,.
\end{align}
\end{subequations}
The remaining components belong to the traceless projection $\hat{\Omega}_{ij|k}$ of the purely spatial tensor $\Omega_{ij|k}$ and it can be dualised into a symmetric rank-two tensor as follows:
\begin{align}
    \pi_{ij} = \pi_{(ij)} := \tfrac12\,\varepsilon_{ikl}\,\hat{\Omega}^{kl|}{}_j \,,
\end{align}
transforming as
\begin{equation}
    \delta \pi_{ij} = - \pd_i \pd_j \tilde \xi_t + \delta_{ij} \pd_k \pd^k \tilde \xi_t \,.
\end{equation}
Substituting everything back in the parent action now leads to
\begin{align} \label{eq:FP-Carroll-magnetic}
\begin{split}
    S_\text{FP-Mag.}^{(c\,\to\,0)}[\tilde{h},\pi] = \int\!\rd t\,\rd^3x \Big[{-\tfrac14}\pd_i\tilde{h}_{jk}\pd^i\tilde{h}^{jk}+\tfrac12\pd^i\tilde{h}_{ij}\pd_k\tilde{h}^{jk}-\tfrac12\pd^i\tilde{h}_{ij}\pd^j\tilde{h}+\tfrac14\pd_i\tilde{h}\pd^i\tilde{h}\hspace{10mm}\\
    {}+\pi^{ij}\Big(\dot{\tilde{h}}_{ij}-2\,\pd_{(i}\tilde{h}_{j)t}\Big) + \tfrac12 \tilde{h}_{tt}\Big(\pd_i\pd^i\tilde{h} - \pd^i\pd^j \tilde{h}_{ij}\Big)\Big]\,,
\end{split}
\end{align}
where $\tilde{h}:=\tilde h^i{}_i$\,.
This is precisely the magnetic theory given in its Hamiltonian form in \cite{Henneaux:2021yzg} (see also \cite{Hansen:2021fxi}).
Therefore, the electric and magnetic Carroll contractions of four-dimensional linearised gravity are dual to each other off-shell, and this equivalence is inherited from the usual electromagnetic duality between the relativistic linearised graviton and dual graviton in four dimensions.

\subsection{Galilean linearised gravity} \label{sec:Fierz-Pauli-Galilei}

\paragraph{Galilean parent action.}

In order to take the Galilean limit of \eqref{eq:parent_Fierz-Pauli_split} we first need to include factors of $c$ in the following way:
\begin{subequations}
\begin{align}
    \pd_0\;&\mapsto\frac1c\,\pd_t\,,&
    \varepsilon^{0ijk}\;&\mapsto\frac1c\,\varepsilon^{ijk}\,,&
    \tilde{e}_{0|}{}^i\;&\mapsto\frac1c\,\tilde{e}_{t|}{}^i\,,&
    \tilde{e}_{i|}{}^0\;&\mapsto c\,\tilde{e}_{i|}{}^t\,,\\
    \Omega_{ij|0}\;&\mapsto\frac1c\,\Omega_{ij|t}\,,&
    \Omega_{0i|j}\;&\mapsto\frac1c\,\Omega_{ti|j}\,,&
    \Omega_{i0|0}\;&\mapsto\frac1{c^2}\,\Omega_{it|t}\,,&
    \Omega_{ij|k}\;&\mapsto\frac1{c^2}\,\Omega_{ij|k}\,.
\end{align}
\end{subequations}
Taking the $c\to\infty$ limit, we find the following parent action for Galilei gravity
\begin{align} \label{eq:parent_FP_Galilei}
\begin{split}
    \hat{S}_\text{FP}^{(c\,\to\,\infty)}\big[\Omega,\tilde{e}\big]
    = \int \rd t\,\rd^3x\,&\Big[{\tfrac14}\,\Omega_{ij|t}\,\Omega^{ij|}{}_t -\Omega_{ij|t}\,\Omega_t{}^{i|j}+\Omega_{t(i|j)}\,\Omega_t{}^{(i|j)} - \Omega_{ti|}{}^i \Omega_{tj|}{}^j\\
    &\hspace{5mm} {} + \varepsilon^{ijk}(2\,\Omega_{it|t}\,\pd_j\tilde{e}_{k|t} + 2\,\Omega_{ti|l}\,\pd_j \tilde{e}_{k|}{}^l - 2\,\Omega_{ij|t}\,\pd_{[t} \tilde{e}_{k]|t})\Big]\,,
\end{split}
\end{align}
with gauge symmetries
\begin{align}
    \delta_\lambda \Omega_{ti|j} &= \pd_i \lambda_j \,,&
    \delta_\lambda \Omega_{ij|t} &= 2\,\pd_{[i}\lambda_{j]} \,,&
    \delta_\lambda \Omega_{it|t} &= - \dot{\lambda}_i \,,
\end{align}
and
\begin{align}
    \delta_{\lambda} \tilde{e}_{i|j} &= \varepsilon_{ij}{}^k \lambda_k \,,&
    \delta_{\tilde{\xi}} \tilde{e}_{i|j} &= \pd_i \tilde{\xi}_j \,,& 
    \delta_{\tilde \xi} \tilde e_{i|t} &= \pd_i \tilde \xi_t \,,& 
    \delta_{\tilde \xi} \tilde e_{t|t} &= \pd_t \tilde \xi_t \,.
\end{align}
This parent action depends neither on $\Omega_{ij|k}$ nor $\tilde{e}_{t|i}$\,, and $\Omega_{it|t}$ is now a Lagrange multiplier.
Moreover, the gauge parameter $\lambda_{ij}$ no longer appears in the linearised gauge transformations and correspondingly Galilean boost symmetry is absent from this system.
This can be expected since the non-linear Galilean theories of \cite{Bergshoeff:2017btm,Hansen:2020pqs} are trivially boost-invariant (they only involve quantities which transform trivially under local Galilean boosts, as opposed to their Carrollian counterparts.)

\paragraph{The electric theory.}

We vary the Galilean parent action with respect to the Lagrange multipliers $\tilde e_{i|j}$\,, $\tilde e_{i|t}$\,, $\tilde e_{t|t}$\,, and $\Omega_{it|t}$\,, obtaining
\begin{align}
    &\Omega_{ti|j} = -\pd_{i}\,e_{t|j} \,,&
    &\Omega_{ij|t} = 2\,\pd_{[i}\,e_{j]|t}\,,&
    &\Omega_{it|t} = 2\,\pd_{[i}\,e_{t]|t}\,,& &\tilde e_{i|t} = \pd_i \tilde \mu_t \,,
\end{align}
up to gauge transformations
\begin{align}
    \delta_\xi e_{a|t} &= \pd_a \xi_t\,,&
    \delta_\xi e_{t|i} &= \xi_i(t)\,,&
    \delta_\lambda e_{t|i} &= - \lambda_i\,,&
    \delta_\lambda e_{i|t} &= \lambda_i\,,&
    \delta \tilde \mu_t &= \tilde \xi_t \,.    
\end{align}
Imposing the gauge $e_{t|i} = 0$ preserved by $\lambda_i = \xi_i(t)$\,, the parent action reduces to
\begin{equation} \label{eq:FP-Galilei-electric}
    S_\text{FP-Elec.}^{(c\,\to\,\infty)}[h] = \int\!\rd t\,\rd^3x\,\pd_{[i} h_{j]t} \pd^{[i} h^{j]}{}_t \,,
\end{equation}
where we noted $h_{it} = h_{ti} = e_{i|t}$\,.
This action corresponds to the linearisation of the leading-order Galilean theory of gravity in \cite{Hansen:2020pqs} around the flat Galilean background given by the background co-metric $h^{\mu\nu} = \text{diag}(0,1,1,1)$ and clock form $\tau_\mu = \delta_\mu^t$\,.
Writing $\tau_\mu = \delta_\mu^t + h_{\mu t}$\,, the linearisation of equation (3.9) in \cite{Hansen:2020pqs} up to quadratic order reads
\begin{equation}
     e\,h^{\mu\nu} h^{\rho\sigma} \pd_{[\mu} \tau_{\rho]} \pd_{[\nu} \tau_{\sigma]} = \pd_{[i} h_{j]t} \pd^{[i} h^{j]}{}_t + \mathcal O(h^3) \,.
\end{equation}
The linearised metric $h_{it}$ transforms\footnote{Note that the parameter of temporal diffeomorphisms $\xi_t$ can be redefined as $\xi_t \mapsto \xi_t - x^i \xi_i(t)$ in order to match with the transformation law of a Galilean clock form under linearised diffeomorphisms.} as $\delta_\xi h_{it} = \pd_i \xi_t + \xi_i(t)$\,.
Note that this action depends neither on $h_{tt}$ nor $h_{ij}$\,, but the full non-linear action for leading-order Galilean gravity does.

\paragraph{The magnetic theory.}

Conversely, varying with respect to $\Omega_{ij|t}$ and $\Omega_{ti|j}$, one finds
\begin{subequations}
\begin{align}
    \frac{\delta \hat S}{\delta \Omega_{ij|t}} = 0 &\quad \Longrightarrow \quad \tfrac12\,\Omega^{ij|}{}_t - \Omega_t{}^{[i|j]} - 2\,\varepsilon^{ijk} \pd_{[t} \tilde{e}_{k]|t} = 0 \,,\\
    \frac{\delta \hat S}{\delta \Omega_{t[i|j]}} = 0 &\quad \Longrightarrow \quad - \Omega^{ij|}{}_t + 2\,\varepsilon^{kl[i}\,\pd_k\tilde{e}_{l|}{}^{j]} = 0 \,,\\
    \frac{\delta \hat S}{\delta \Omega_{t(i|j)}} = 0 &\quad \Longrightarrow \quad \Omega_t{}^{(i|j)} - \delta^{ij}\,\Omega_{tk|}{}^k + \varepsilon^{kl(i}\,\pd_k\tilde{e}_{l|}{}^{j)} = 0 \,,
\end{align}
\end{subequations}
which gives, upon replacement in the parent action and imposing the gauge $\tilde{e}_{[i|j]} = 0$
\begin{align} \label{eq:FP-Galilei-magnetic}
\begin{split}
    S_\text{FP-Mag.}^{(c\,\to\,\infty)}\big[\tilde{h},A\big] =\int\!\rd t \, \rd^3 x\,\Big[{-\tfrac14}\pd_i\tilde{h}_{jk}\pd^i\tilde{h}^{jk}+\tfrac12\pd^i\tilde{h}_{ij}\pd_k\tilde{h}^{jk}-\tfrac12\pd^i\tilde{h}_{ij}\pd^j\tilde{h}+\tfrac14\pd_i\tilde{h}\pd^i\tilde{h}\hspace{15mm}\\
    {} + \tfrac12\,\tilde{h}_{tt} \left(\pd_i\pd^i\tilde{h} - \pd^i\pd^j \tilde{h}_{ij} \right) + \tilde{h}_{it} \left( \pd_j\dot{\tilde{h}}^{ij} - \pd^i \dot{\tilde{h}}\,\right) + A^{ij}\,\pd_{[i} \tilde{h}_{j]t} \Big]\,,
\end{split}
\end{align}
where $\tilde{h}:=\tilde{h}^i{}_i$\,.
Here we have identified $\tilde{h}_{ij} = 2\,\tilde{e}_{(i|j)}$\,, $\tilde{h}_{ti} = \tilde{h}_{it} = \tilde{e}_{i|t}$\,, and $\tilde{h}_{tt} = 2\,\tilde{e}_{t|t}$\,, and we have defined
\begin{align}
    A^{ij} = A^{[ij]} &:= 2\,\varepsilon^{ijk} \Omega_{kt|t} \,,&
    \delta A^{ij} &= 2\,\pd^{[i} \dot{\tilde \xi}^{j]} \,.
\end{align}
Notice that $\delta_{\tilde{\xi}} \tilde{h}_{it} = \delta_{\tilde{\xi}} \tilde{h}_{ti} = \pd_i \tilde{\xi}_t$\,, as expected for a Galilean metric. 
This theory corresponds to the transformation of the next-to-leading order theory of \cite{Hansen:2020pqs} by introducing the auxiliary field $A^{ij}$ before taking the limit $c \to \infty$.
This theory was actually first introduced in \cite{Bergshoeff:2017btm} by gauging the Galilei algebra and it can be seen that the Lagrange multiplier $A^{ij}=A^{[ij]}$ enforces the linearisation of the twistless torsional condition (TTNC) $\tilde \tau \wedge \rd \tilde \tau = 0$\,.
Indeed, writing $\tilde \tau_\mu = \delta_\mu^t + \tilde h_{\mu t}$\,, we have
\begin{equation}
     \tilde \tau \wedge \rd \tilde \tau = \pd_{[i} \tilde h_{j] t} \, \rd t \wedge \rd x^i \wedge \rd x^j + \mathcal O(\tilde h^2) \,.
\end{equation}

This contraction, referred to as Galilei gravity in \cite{Bergshoeff:2017btm,Hansen:2020pqs}, coincides with the massless limit of Newton-Cartan gravity.
As a reminder, Newton-Cartan gravity \cite{Cartan:1923zea,Cartan:1924yea} is a reformulation of Newtonian gravity along similar geometric grounds as general relativity and can be obtained by gauging the Bargmann\footnote{Note that in order to recover Newton gravity, some equations of motion still have to be imposed by hand. By contrast, a completely off-shell formulation was proposed in \cite{Hansen:2019pkl} based on the gauging of a bigger algebra.} algebra \cite{Andringa:2010it} which is the unique central extension of the Galilei algebra by a generator interpreted as the mass so that the massless limit is equivalent to a quotient.

\section{Conclusion} \label{sec:conclusions}

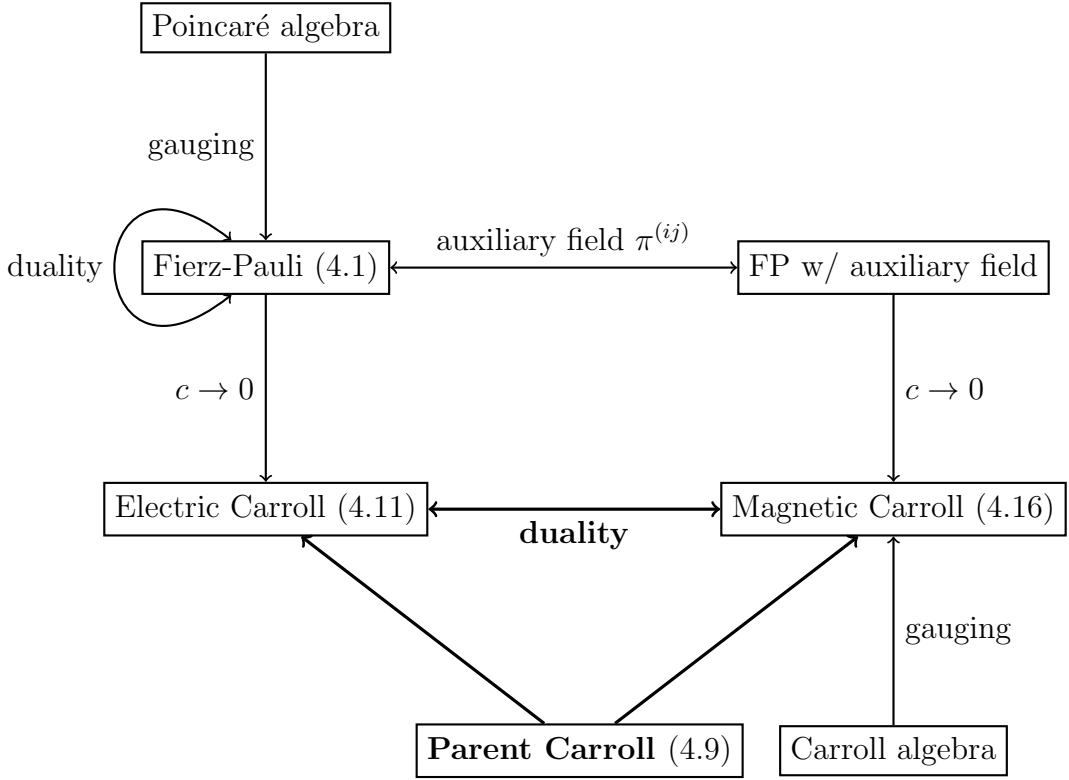
\begin{figure}[h]
    \centering
    \begin{tikzpicture}[node/.style={rectangle, draw=black, thick, minimum size=5mm}, every node/.style=midway]
  \matrix[column sep={10em,between origins}, row sep={6em}] at (0,0) {
    \node[node](Po) {Poincar\'e algebra} ; && \\
    \node[node](FP) {Fierz-Pauli \eqref{eq:FP}} ; && \node[node](FPX) {FP w/ auxiliary field}; \\
    \node[node](El) {Electric Carroll \eqref{eq:FP-Carroll-electric}}; && \node[node] (Ma) {Magnetic Carroll \eqref{eq:FP-Carroll-magnetic}}; \\
    & \node[node] (Pa) {\textbf{Parent Carroll \eqref{eq:parent_FP_Carroll}}}; & \node[node] (Ca) {Carroll algebra} ; \\
  };
  \draw[->, thick] (Po) -- (FP) node[anchor=east] {gauging};
  \draw[->, thick] (FP) -- (El) node[anchor=east]  {$c \to 0$};
  \draw[<->, thick] (FP) .. controls +(-2.5,-2) and +(-2.5,+2) .. (FP) node[anchor=east] {duality};
  \draw[<->, thick] (FP) -- (FPX) node[anchor=south] {auxiliary field $\pi^{(ij)}$};
  \draw[->, thick] (FPX) -- (Ma) node[anchor=west] {$c \to 0$};
  \draw[<->, very thick] (El) -- (Ma) node[anchor=north] {\textbf{duality}};
  \draw[->, thick] (Ca) -- (Ma) node[anchor=west] {gauging};
  \draw[->, very thick] (Pa) -- (El) ;
  \draw[->, very thick] (Pa) -- (Ma) ;
\end{tikzpicture}
    \caption{Relation between the different Carrollian limits of linearised general relativity.}
    \label{fig:Carroll-FP}
\end{figure}

\begin{figure}[h]
    \centering
    \begin{tikzpicture}[node/.style={rectangle, draw=black, thick, minimum size=5mm}, every node/.style=midway]
  \matrix[column sep={10em,between origins}, row sep={6em}] at (0,0) {
    \node[node](Po) {Poincar\'e algebra} ; && \\
    \node[node](FP) {Fierz-Pauli \eqref{eq:FP}} ; && \node[node](FPX) {FP w/ auxiliary field}; \\
    \node[node](El) {Electric Galilei \eqref{eq:FP-Galilei-electric}}; && \node[node] (Ma) {Magnetic Galilei \eqref{eq:FP-Galilei-magnetic}}; \\
    & \node[node] (Pa) {\textbf{Parent Galilei \eqref{eq:parent_FP_Galilei}}}; & \node[node] (Ga) {Galilei algebra} ; \\
  };
  \draw[->, thick] (Po) -- (FP) node[anchor=east] {gauging};
  \draw[->, thick] (FP) -- (El) node[anchor=east]  {$c \to \infty$};
  \draw[<->, thick] (FP) .. controls +(-2.5,-2) and +(-2.5,+2) .. (FP) node[anchor=east] {duality};
  \draw[<->, thick] (FP) -- (FPX) node[anchor=south] {auxiliary field $A^{[ij]}$};
  \draw[->, thick] (FPX) -- (Ma) node[anchor=west] {$c \to \infty$};
  \draw[<->, very thick] (El) -- (Ma) node[anchor=north] {\textbf{duality}};
  \draw[->, thick] (Ga) -- (Ma) node[anchor=west] {gauging};
  \draw[->, very thick] (Pa) -- (El) ;
  \draw[->, very thick] (Pa) -- (Ma) ;
\end{tikzpicture}
    \caption{Relation between the different Galilean limits of linearised general relativity.}
    \label{fig:Galilei-FP}
\end{figure}
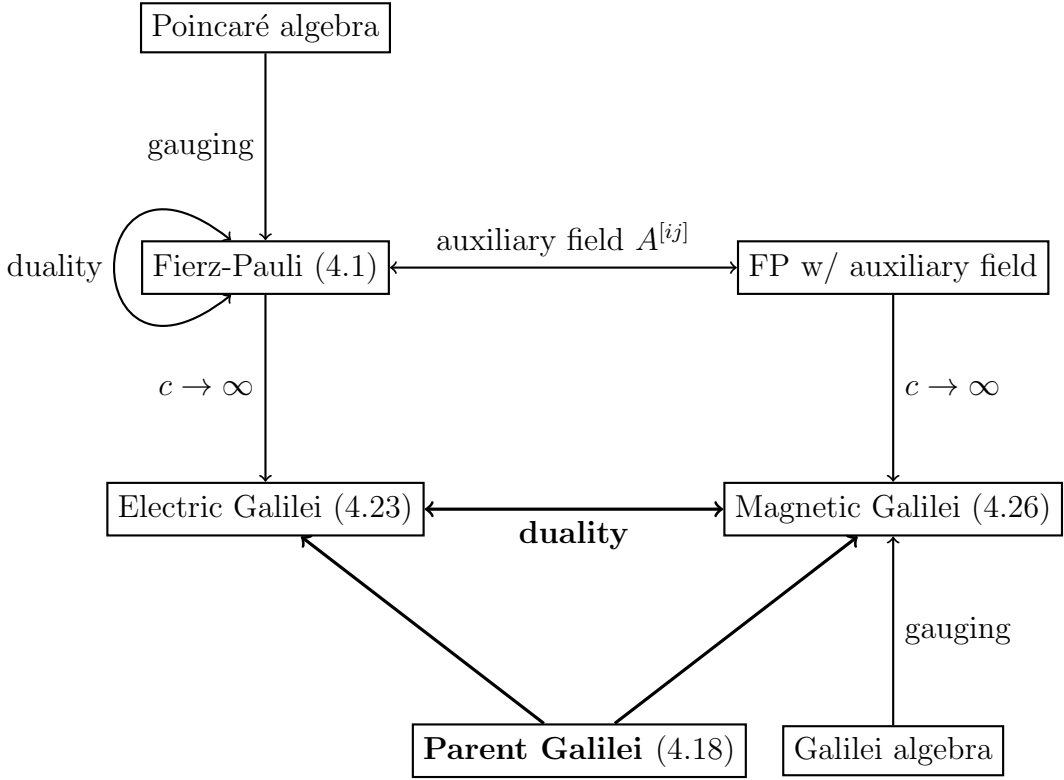

In this paper we have dualised several Carrollian and Galilean theories which were previously obtained by taking limits of relativistic electromagnetism and linearised gravity.
Our off-shell dualisation procedure, which sometimes relates two identical theories in the relativistic case, relates the two possible non-Lorentzian limits that were identified in the literature, both in the Carrollian and in the Galilean contractions.
The electric and magnetic limits of Maxwell theory in four dimensions were already known to be dual to each other on-shell, but here we have extended this off-shell by showing that their respective action principles are indeed dual to each other.
We found a natural generalisation of this result to $p$-forms in arbitrary dimensions, including scalar fields.
In the case of linearised Carrollian gravity, we have found an analogous duality between the electric and magnetic theories, thus showing that they are equivalent.
Similarly, we found that the leading order and next-to-leading order theories in the large-$c$ expansion of Fierz-Pauli are related to each other with another such duality symmetry.
As a remark, the electric Carrollian (resp. Galilean) limit of general relativity can be obtained as the $c \to 0$ (resp. $c \to \infty$) limit of the Einstein-Hilbert action, while their magnetic counterparts can be obtained either by taking a limit of the Einstein-Hilbert action after introducing an auxiliary field or by gauging the corresponding symmetry algebra \textit{\`a la} Cartan.
Although the electric and magnetic theories look very different, there exist non-local transformations mapping one into the other.
It should also be possible to check that both theories feature the same solution space and the same number of degrees of freedom along the lines of \cite{Boulanger:2020yib}, although we leave this for future exploration. We summarise the uncovered duality symmetries for linearised gravity in four dimensions in Figures~\ref{fig:Carroll-FP} and \ref{fig:Galilei-FP}.

One can build upon these results in a number of ways. For instance, one can try to dualise free higher-spin fields \cite{Hull:2001iu,Boulanger:2003vs,Bergshoeff:2016soe} or mixed-symmetry fields \cite{Boulanger:2012df,Boulanger:2015mka,Henneaux:2019zod}.
In particular, linearised gravity in $D>4$ dimensions is dual to a mixed-symmetry field called the dual graviton, and the same is expected to hold for their Carrollian and Galilean counterparts.
One can also look at higher dualisations where a field is dualised over the indices of another, possibly empty, column in its Young tableau \cite{Boulanger:2012df,Boulanger:2015mka,Boulanger:2022arw}.

Both the electric and magnetic branches of Carrollian conformal field theories are known to be relevant in the context of flat-space holography \cite{Mason:2023mti,Alday:2024yyj}.
However, they admit different classes of deformations, different quantum aspects \cite{deBoer:2023fnj} and couple differently to Carrollian background geometry, producing different stress tensors \cite{deBoer:2021jej,Rivera-Betancour:2022lkc,Baiguera:2022lsw}.
This may indicate that electric and magnetic Carrollian field theories play different, possibly complementary, roles in holography or they may be associated with different notions of holography altogether.
For example, it was found in \cite{Donnay:2022wvx} that the putative Carrollian field theory dual to gravity in four-dimensional asymptotically flat spacetimes was of electric type.
In contrast, it was found in \cite{Barnich:2012rz} that the `magnetic' limit of Liouville theory admits a deformation giving rise to a central extension in the BMS$_3$ charge algebra, as opposed to the `electric' limit.
This central extension can be matched with the Brown-Henneaux central charge \cite{Brown:1986nw}, signaling that this two-dimensional field theory is able to capture the boundary dynamics of bulk three-dimensional gravity for asymptotically flat space-times.
However, since there exist non-local duality symmetries between the two-dimensional free electric and magnetic Carrollian scalar field theories, their respective holographic duals should still be related by (possibly elaborate) transformations in the bulk.

Another interesting avenue would be to work out the effect of a Carrollian electromagnetic duality of the boundary field theory on the dynamics of gravity in the bulk in a more general context.
This analysis was started recently in \cite{Mittal:2022ywl}, where it was shown that the Ehlers symmetry \cite{Ehlers:1957} in the bulk of four-dimensional asymptotically locally flat space-times directly gives rise to an electromagnetic duality on the Carrollian boundary relating the boundary Carrollian stress tensor to the boundary Carrollian Cotton tensor.
In view of the growing interest in Carrollian field theories for flat holography it would be interesting explore this in more detail.

One direction could be to enlarge Ehlers in the bulk to include larger duality symmetries such as Geroch \cite{Geroch:1970nt,Geroch:1972yt,Bossard:2017aae,Bossard:2018utw,Bossard:2021jix,Bossard:2024gry}, hyperbolic BKL \cite{Belinsky:1982pk,Nicolai:1991kx,Damour:2002et,Damour:2002cu,Nicolai:2005su,Cederwall:2021ymp}, and `very-extended' \cite{Kleinschmidt:2003mf,Riccioni:2007hm,Tumanov:2015yjd,Tumanov:2016abm,Bossard:2017wxl,Bossard:2019ksx,Bossard:2021ebg,Glennon:2020qpt,Glennon:2024ict} symmetries, and to study their effects on the boundary.
In another direction, the tenacious reader might try to work out non-Lorentzian contractions of relativistic field theories with manifest extended Ehlers symmetries \cite{Hohm:2013pua,Majumdar:2019vic}.
Such `extended' (double, exceptional, etc.) field theories are formulated in terms of generalised space-times, and understanding their non-Lorentzian contractions would be a first step towards understanding extended non-Lorentzian duality symmetries.

It would also be worth investigating other types of duality than the ones considered here.
For instance, it may very well be possible to work out a possible duality between electric and magnetic $p$-forms, rather than between, say, an electric $p$-form and magnetic \mbox{$(D-p-2)$-form}.
We have also not considered the possible Carroll-Galilei dualities along the lines of \cite{Duval:2014uoa,Figueroa-OFarrill:2022pus} but they deserve further attention.
We have only focused here on free theories as there are long-standing problems associated with off-shell dualisation for interacting theories.
Non-linear `ModMax' electrodynamics admits Galilean and Carrollian limits that were studied in \cite{Banerjee:2022sza,Chen:2024vho}.
These theories respect the duality symmetry of the non-interacting theory and it would be interesting to see if such theories also admit a parent-action description.
A parent action relating non-linear gravity and its dual was also put forward in \cite{Boulanger:2008nd}, but it features more fields than the ones needed to describe the parent action for linearised gravity, and this leads to a more challenging non-Lorentzian analysis.
It is also worth mentioning that duality symmetries can also exist on backgrounds with non-zero scalar curvature as demonstrated, for example, in \cite{Julia:2005ze}.
Let us conclude by commenting that this work fits into the long-term goal of building non-Lorentzian analogues of supergravity and holography, and our results may serve as a useful tool in order to build such theories and uncover new dualities.

\section*{Acknowledgements}

We wish to thank the Erwin Schr\"odinger Institute in Vienna for hospitality during the workshop `Carrollian Physics and Holography' where this work was initiated. SP also thanks the Aristotle University of Thessaloniki where part of these results were presented.
SP wishes to thank Glenn~Barnich and Tasos~Petkou for interesting discussions and suggestions.
JAO is supported by the \emph{Fonds National de la Recherche Scientifique} (FNRS), grant number FC 43791.
SP is supported by the \emph{Fonds Friedmann} run by the \emph{Fondation de l'\'Ecole polytechnique}.

\addcontentsline{toc}{section}{References}
\bibliographystyle{utphys}

\end{document}